\documentclass[a4paper]{article}
\usepackage{INTERSPEECH2022}
\usepackage{multirow}
\usepackage{cite}

\title{Low-data? No problem: low-resource, language-agnostic conversational text-to-speech via F0-conditioned data augmentation}

\name{Giulia Comini, Goeric Huybrechts, Manuel Sam Ribeiro, Adam Gabryś, Jaime Lorenzo-Trueba}
\address{Amazon Alexa, TTS Research}
\email{
    \{gcomini, huybrech, manuerib, gabrysa, truebaj\}@amazon.com}

\begin{document}
%
\maketitle
\begin{abstract}
The availability of data in expressive styles across languages is limited, and recording sessions are costly and time consuming. To overcome these issues, we demonstrate how to build low-resource, neural text-to-speech (TTS) voices with only 1 hour of conversational speech, when no other conversational data are available in the same language. 
Assuming the availability of non-expressive speech data in that language, we propose a 3-step technology: 1) we train an F0-conditioned voice conversion (VC) model as data augmentation technique; 2) we train an F0 predictor to control the conversational flavour of the voice-converted synthetic data; 3) we train a TTS system that consumes the augmented data. We prove that our technology enables F0 controllability, is scalable across speakers and languages and is competitive in terms of naturalness over a state-of-the-art baseline model, another augmented method which does not make use of F0 information.

\end{abstract}
\noindent\textbf{Index Terms}:
text-to-speech, expressive speaking styles, F0 prediction, voice conversion, data augmentation

\section{Introduction}

Neural text-to-speech (TTS) has been shown to produce high-quality synthetised speech from written text. 
State-of-the-art TTS models usually require a large amount of high-quality speech data \cite{oord2016wavenet, oord2017parallel, Sotelo2017Char2WavES, arik2017deep, shen2018natural, ping2019clarinet, li2019neural, ren2021fastspeech}. Traditionally, these data were recorded in a flat, neutral style, however, with the advancement in TTS research, a new interest in various speech styles has arisen \cite{wang2018style, skerry2018towards, lee2021styler, huybrechts2021low, shah2021non}.
Thus, in order to generate a varied portfolio of TTS voices which covers voices in multiple styles and languages, a large quantity of data is needed. 
Consequently, in an effort to mitigate data collection, which is costly and time consuming, interest in low-resource TTS research has grown significantly.
The two main research trends of low-resource TTS focus on: 1) transfer learning to a low-resource voice, leveraging a large multi-speaker dataset \cite{arik2017deep2, ping2018deep, taigman2018voiceloop, arik2018neural, lee2018voice, jia2019transfer, tu2019endtoend, tits2019exploring, zhang2020unsupervised, cooper2020zeroshot, casanova2021scglowtts}; 2) data augmentation, leveraging multi-speaker data \cite{xu2020lrspeech}, voice conversion (VC) capabilities \cite{huybrechts2021low, shah2021non}, and TTS \cite{HwangYSK21}. The main limitation of these approaches is the assumption 
that a large amount of data are available for any language and style. In order to make the technology scalable, we propose a language-agnostic method which relies on a minimal amount of such data, regardless of its style.
Our work is inspired by \cite{huybrechts2021low}, which shows that it is possible to build high-quality voices with only a small amount of speech data (\emph{target data}). \cite{huybrechts2021low} leverages an unlimited amount of data from other voices (\emph{supporting data}), which are converted to the target voice via VC. A multi-speaker TTS model is then trained with the original and augmented data. 
We remove the main constraint of \cite{huybrechts2021low}, i.e. the assumption that a large amount of expressive data are available for any language; we assume instead to have flat, inexpressive, \emph{neutral-style} supporting data. We also reduce the amount of supporting data from an ``unlimited amount" to 8-10 hours. Further, our low-resource target data are recorded in an expressive, friendly and spontaneous style, which we call \emph{conversational style}. 
We show that our technology is language-agnostic, building 9 voices in 5 locales, while using the same configuration of hyper-parameters.
As in \cite{huybrechts2021low} we perform data augmentation via VC, using a model based on Copycat \cite{karlapati2020copycat}. Since we have supporting neutral data, but we would like the augmented data to retain the conversational style of the target voice, we enable the possibility of controlling the output speaker style during VC, leveraging log-F0 (F0) information. Controlling prosody with F0 information has been extensively explored in both VC \cite{qian2020f0} (especially for emotion conversion \cite{emoDualVC2019, zhou2020transforming, zhouEmoST2021, veillon2021endtoend}) and TTS literature \cite{arik2017deep, arik2017deep2, ren2021fastspeech, 2021fastpitch, mellotron2020}.
We learn how to model F0 through a separate model, which is used to produce F0 trajectories in the expressive, personalised style of the target speaker. The output of such model is then used to feed the VC model at inference time.
Our 3-steps pipeline consists of: 1) an F0-conditioned VC model to do data augmentation; 2) an F0 predictor that controls the conversational style of the voice-converted synthetic data; 3) a TTS system that consumes the augmented data.
The main contributions of this paper are: 1) an F0 predictor model that produces reliable F0 trajectories; 2) a method to control the style of the augmented samples consumed by the TTS model; 3) a scalable, language-agnostic technology for building high-quality, low-resource TTS voices.



\section{Methodology}
\label{sec:methodology}
Our goal is to build low-resource, neural TTS voices with only 1 hour of conversational speech, when no other conversational data are available in the same language. In order to do so, we additionally require 8-10 hours of neutral supporting data from another voice in the same gender and locale as the target voice. Given these conditions, we propose a 3-step technology. Firstly, a modified version of Copycat \cite{karlapati2020copycat} (VC) which makes use of F0 trajectories is trained in order to augment the target speaker data. Secondly, a neural F0 predictor learns to predict F0 trajectories, which are in turn injected into the VC system at inference time. In this way we leverage the supporting data, which are converted to the target speaker identity, but also the target data, from which we learn the conversational style that we want to maintain during VC. Finally, a single speaker TTS model is trained with the original and synthetic data.

\subsection{F0-conditioned Voice Conversion}
\label{subsec:vc}

\begin{figure}[t]
  \centering
  \includegraphics[width=1.01\linewidth]{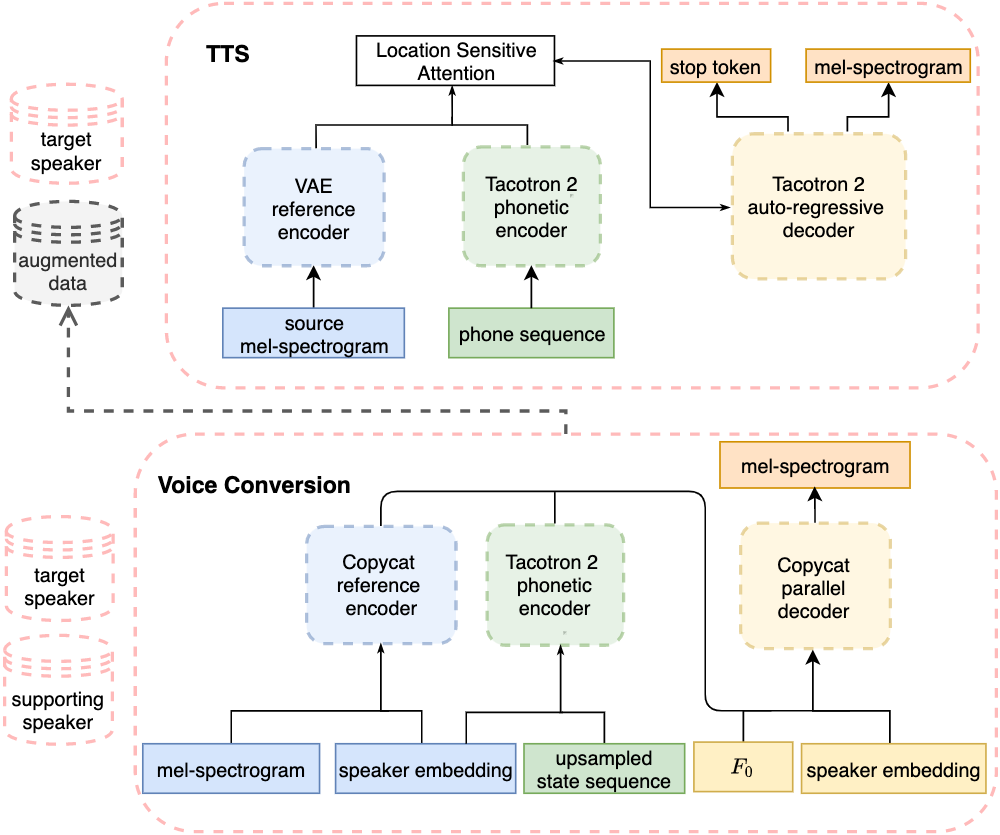}
  \caption{Voice Conversion + TTS pipeline}\vspace{-1mm}
  \label{fig:vc_model}
\end{figure}


Our proposed VC approach (Figure  \ref{fig:vc_model}) is a modification of CopyCat \cite{karlapati2020copycat}, with two additions: the concatenation of the speaker embeddings to the upsampled phonemes (before passing them to the phoneme encoder), as in \cite{huybrechts2021low}; frame-level F0 as input to the decoder, similarly to \cite{qian2020f0}, except for the fact that we do not apply normalisation. The idea behind this architecture is to learn to disentangle between linguistic information, prosody content and speaker identity, so that VC is performed effectively.
\cite{karlapati2020copycat} uses a variational auto-encoder (VAE) \cite{kingma2014autoencoding} to preserve the supporting data prosody, downsampling and then upsampling the latent spectrogram representation along the time dimension. If the supporting speaker is recorded in a neutral, inexpressive style, her prosody will be passed onto the target speaker during inference. However, in our use-case we would like to maintain the conversational, spontaneous style of the target speaker.
To be able to control the style of the converted samples at inference time, we condition the decoder with F0, interpolating it over the unvoiced phonemes. Any other prosodic information which is not F0 and duration is still going to be carried over by the reference encoder. 
The model is trained on the target and supporting speakers’ data for 100k steps with batch size of 32, and then fine-tuned on the target speaker data for an additional 4k steps. Fine-tuning has been shown to remove the potential remaining speaker leakage \cite{huybrechts2021low}, without degrading the quality of the synthetic samples. F0 is extracted by the WORLD vocoder \cite{Morise2016WORLDAV}. We use a Kullback–Leibler divergence (KLD) loss for training the VAE reference encoder and an L1 loss is computed between the oracle and the predicted mel-spectrogram.

\begin{figure}
  \centering
  \includegraphics[width=0.9\linewidth]{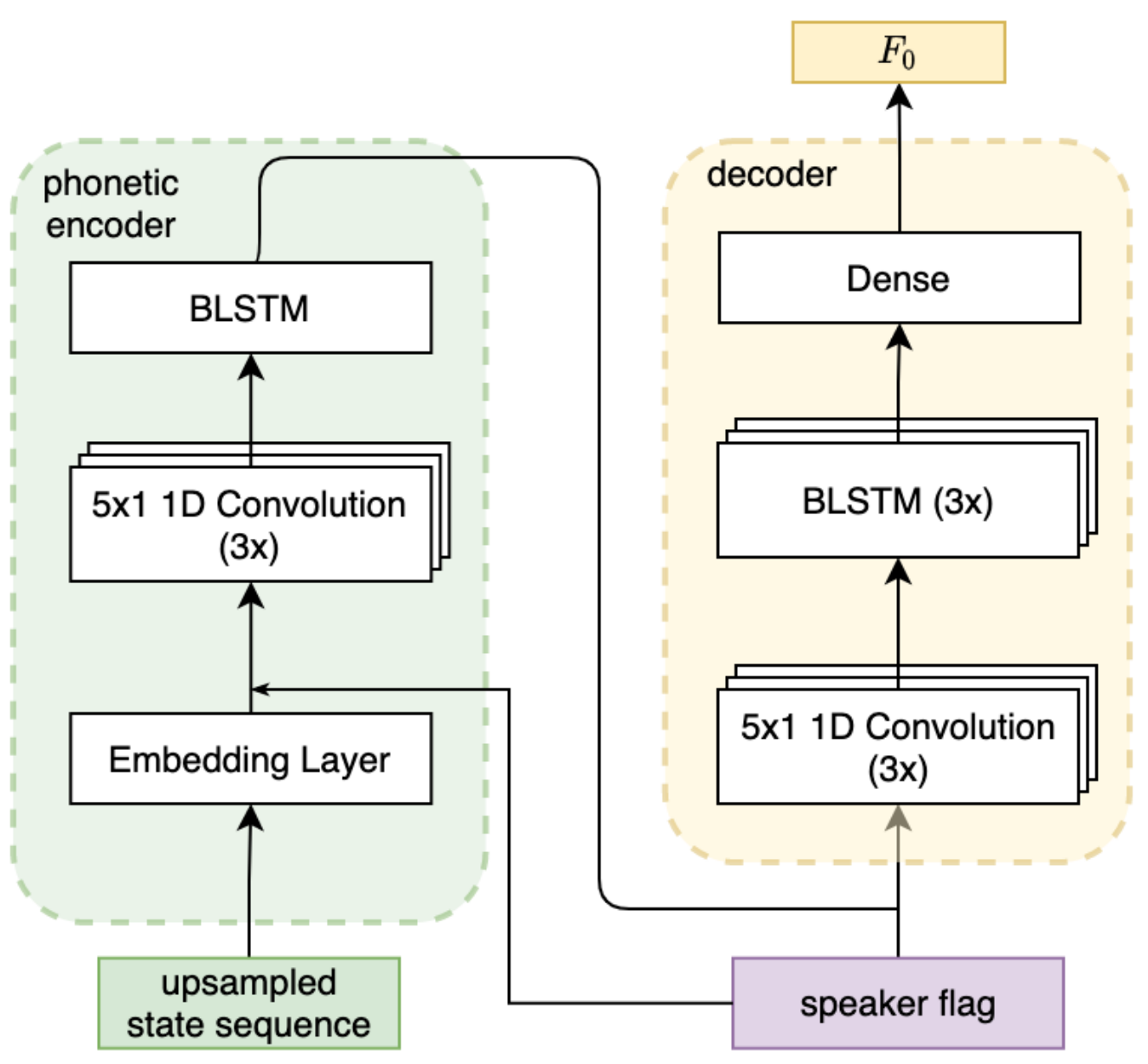}
  \caption{F0 predictor model architecture}\vspace{-1mm}
  \label{fig:f0_predictor}
\end{figure}

\subsection{F0 predictor}
\label{subsec:f0_pred}
As mentioned in the previous section, our aim is to generate synthetic data which mimic the conversational style of the target speaker voice. The architecture of the VC system (Section \ref{subsec:vc}) allows us to control prosody at inference time, conditioning the decoder with an F0 trajectory. Since we do not want to maintain the neutral-style F0 trajectories of the supporting speaker, we learn to reproduce the target speaker F0 by training a separate F0 predictor. The F0 predictor output is then used in the VC system at inference time. The model architecture (Figure \ref{fig:f0_predictor}) consists of: a Tacotron 2 text encoder \cite{shen2018natural} which, similarly to the VC system, receives as input state-level phonetic information upsampled at frame level; a parallel decoder that, given the encoded phonemes and a one-hot speaker ID flag, is trained to predict the F0 trajectories. The F0 predictor is trained for 10k steps on the neutral supporting and conversational target speaker data, and fine-tuned for another 10k steps on the target speaker data only. Interpolation of F0 is used for assigning an F0 value to the unvoiced phonemes. An L1 loss is computed between the oracle and the predicted F0 trajectories. 

\subsection{Text-to-Speech}
\label{subsec:tts}

Our TTS model is based on Tacotron 2 \cite{shen2018natural}. As highlighted in Figure \ref{fig:vc_model}, it consists of a phonetic encoder, an additional VAE \cite{kingma2014autoencoding} that captures prosodic information \cite{zhang2019learning, tyagi2020}, and an auto-regressive decoder. A single-head location-attention mechanism \cite{bahdanau2014neural} is used between the encoder and the decoder. Our TTS systems are trained on 1 hour of target speaker real data and 8-10 hours of synthetic data, generated by the F0-conditioned VC system (where the F0 trajectory is the one predicted by the F0 predictor of Section \ref{subsec:f0_pred}). As in \cite{huybrechts2021low}, we additionally fine-tune the TTS models for 4k steps on the target speaker data. Fine-tuning guides the model towards the space of the real data, which results in a refinement in signal and segmental quality, as well as expressiveness. At inference time we condition the decoder with a VAE z-vector computed averaging the target speaker real data. Finally, the predicted mel-spectrograms are converted to audio waveforms by a Parallel Wavenet universal neural vocoder \cite{jiao2021universal}.


\section{Experiments}

In this section we assess the strengths of our approach, performing objective analyses of the components of the data augmentation pipeline, as well as perceptual evaluations on the final TTS voices. More specifically, we aim to demonstrate that: 1) the F0-predictor model produces conversational F0 trajectories; 2) we are able to achieve a good level of F0 controllability in VC; 3) the voice-converted samples resemble the target speaker F0 distribution; 4) our proposed technology is able to build low-resource TTS voices and either improves or is on par with the previous best augmented system \cite{huybrechts2021low} across different languages.

\textbf{Experimental setting.}
To demonstrate that our technology is scalable in terms of languages and voices, we build 9 conversational TTS voices\footnote{Samples available on www.amazon.science/low-data-no-problem}, of which 5 are females and 4 are males, in 5 different locales (Canadian French, French, Italian, German and Spanish), using only 1 hour of data for each target voice. In each experiment we additionally use 8-10 hours of supporting speaker data, i.e. neutral recordings from a voice of the same gender as the target speaker. The data used in these experiments have been professionally recorded and are part of our internal dataset. Every model has been trained in parallel on 4 GPUs (Tesla V100-SXM2-32GB) with batch size of 32, and the same hyper-parameters have been used across all the experiments. The test set consists of 200 samples per voice. For each target voice we also build a baseline system based on \cite{huybrechts2021low}, which is trained in the same fashion as the proposed one, with the only difference that the synthetic data have been generated with a VC conversion system that does not make use of F0 (therefore the F0-predictor is not needed). Everything else besides the F0-conditioning component is as described in Sections \ref{subsec:vc} and \ref{subsec:tts}. 

\begin{table*}[t!]
\caption{F0 distribution analysis. We highlight the differences in terms of F0 mean and variance between the target speakers and their respective supporting speaker, since it has an impact on the models' performances. The KLD values are computed between the F0/delta-F0 distributions of synthetic samples with respect to the target data F0/delta-F0 distributions. Values closer to 0 are preferred.}\vspace{-1mm}

\begin{tabular}{ccccccc}
\hline
 &
\multirow{2}{*}{\textbf{Difference in F0 mean}} &
\multirow{2}{*}{\textbf{Difference in F0 variance}} &
\multicolumn{3}{c}{\textbf{KLD}} \\
 &
\multirow{2}{*}{\textbf{(wrt Target)}} &
\multirow{2}{*}{\textbf{(wrt Target)}} &
\multicolumn{3}{c}{\textbf{(wrt Target F0 $\vert$ delta-F0 distributions)}} \\
 &
 &
 &
\textbf{No-F0 VC} &
\textbf{Orig-F0 VC} &
\textbf{Pred-F0 VC} \\
\hline
\textbf{Speaker 1 (it)} & 
+47.6 & 
+1840.1 & 
0.27 $\vert$ \textbf{0.02} & 
0.64 $\vert$ \textbf{0.02} & 
\textbf{0.10} $\vert$ \textbf{0.02} \\
\textbf{Speaker 2 (fr-ca)} & 
+31.4 & 
+1637.8 & 
\textbf{0.05} $\vert$ \textbf{0.01} & 
0.10 $\vert$ 0.02 & 
0.08 $\vert$ \textbf{0.01} \\
\textbf{Speaker 3 (fr)} & 
-9.27 & 
+980.39 & 
\textbf{0.06} $\vert$ 0.04 & 
0.32 $\vert$ 0.08 & 
\textbf{0.06} $\vert$ \textbf{0.03} \\
\textbf{Speaker 4 (es)} & 
+29.7 & 
+484.6 & 
0.09 $\vert$ 0.02 & 
\textbf{0.01} $\vert$ \textbf{0.01} & 
0.06 $\vert$ 0.02 \\
\hline
\end{tabular}
\label{table:f0_distribution_analysis}
\end{table*}

\subsection{Objective analysis}
\textbf{F0 predictor quality assessment.}
Following the procedure described in Section \ref{subsec:f0_pred}, we train an F0 predictor for each one of the 9 target voices. Every model is trained on target and supporting speaker pairs. 
We use the Root Mean Square Error (RMSE) and Pearson’s Correlation Coefficient (Correlation) as metrics to compare the predicted F0 trajectories and the oracle ones in the log domain. The objective metrics are computed on an interpolated F0. 
Across all the voices, we get an average RMSE of 49.6 and 0.65 for Correlation. We will explain the implication of the results later in this section

\begin{figure}[t]
  \centering
  \includegraphics[width=1.0\linewidth]{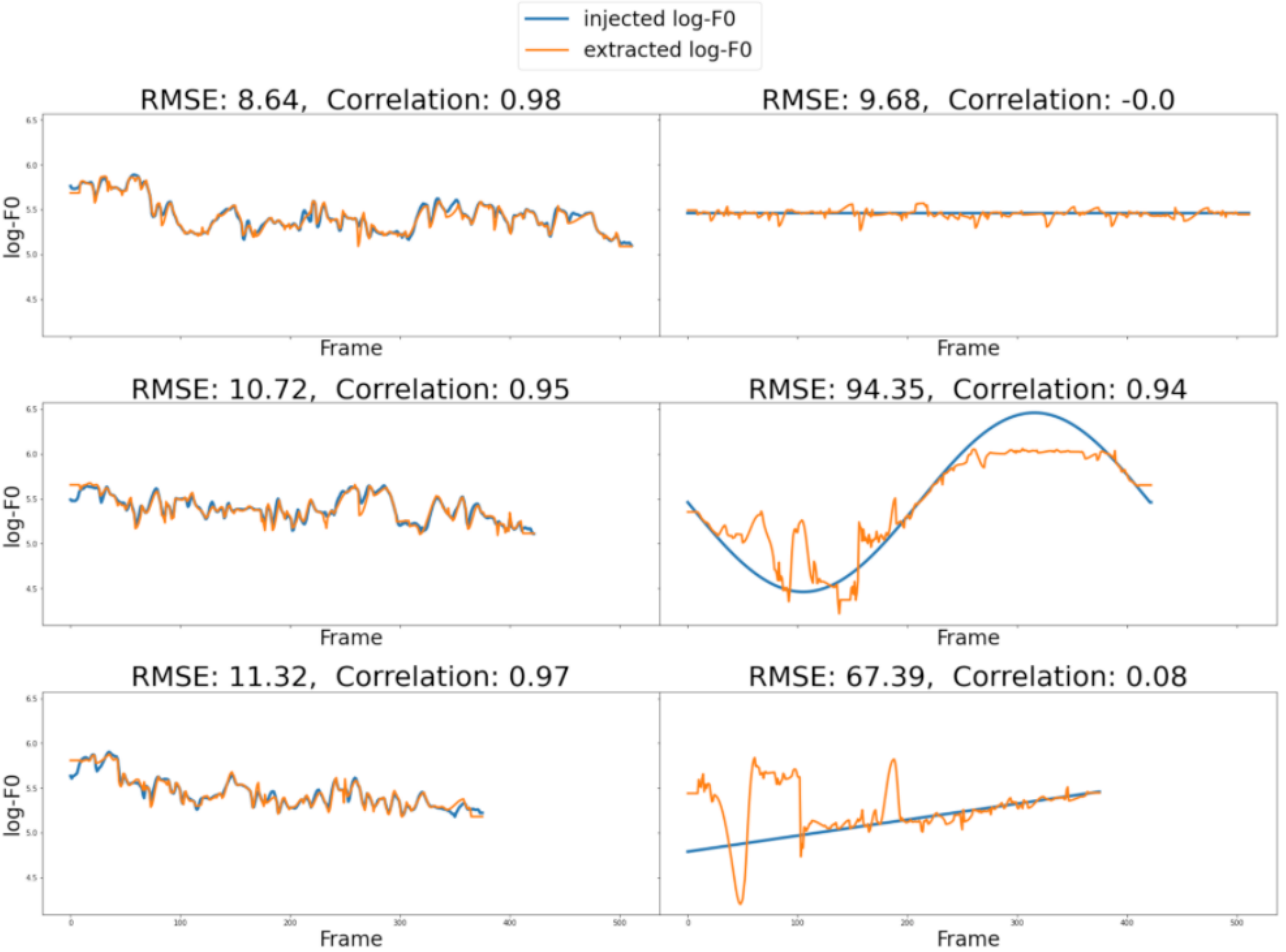}
  \caption{F0 controllability. The blue lines correspond to the F0 trajectories used to condition the decoder in VC, while the orange ones are extracted from the synthetic samples.}\vspace{-2mm}
  \label{fig:f0_controllability}
\end{figure}

\textbf{F0 controllability in VC.}
The main advantage of an F0-conditioned model is the ability to gain control over F0 at inference time, which is crucial for handling speaking styles. In this analysis we show how the proposed VC system (Section \ref{subsec:vc}) is able to follow the injected F0 trajectory. The left column of Figure \ref{fig:f0_controllability} presents three examples of injected trajectories (in blue) together with their counterpart extracted from synthetic audio files (in orange). The injected trajectories come from the F0 predictor model from Section \ref{subsec:f0_pred}. In the second column of Figure \ref{fig:f0_controllability}, we show how, for the same source samples of the left column, the extracted F0 changes when we inject artificially constructed trajectories. Our model does well in following realistic F0s, while it struggles with the artificial trajectories. It can be argued that the decoder is trained to reconstruct somewhat ``intelligible" speech, while the artificial trajectories might not follow the short-term F0 variation that depends on the phonetic sequence. The result is essentially a compromise between an attempt to both reconstruct speech and follow the F0 trajectory.
Over 200 converted samples, the average RMSE between the injected and extracted F0 curves is 12.01 and the average Correlation is 0.94. As a reference, we compute the same values when injecting the flat line, the sinusoidal curve and the linear ascendent line, obtaining, as average over the 200 samples, 13.56, 106.42 and 76.87 for RMSE, and 0, 0.86 and 0 for average Correlation, respectively. Note that from a mathematical perspective, a correlation of 0 between the flat linear trajectories and the extracted ones is expected.

\begin{table*}[t]
\caption{Results for the Naturalness MUSHRA evaluation. TTS from No-F0 VC is the baseline system; TTS from F0 VC is the proposed system; the p-value comes from the Student's t-test and it is computed between the baseline and the proposed system; the \% of improvement is the relative improvement of the proposed system with respect to the baseline and the recordings.}\vspace{-1mm}
\label{table:mushra}
\begin{tabular}{ccccccc}
\hline
\multirow{2}{*}{\textbf{Locale}} &
  \multirow{2}{*}{\textbf{Gender}} &
  \multirow{2}{*}{\textbf{Recordings}} &
  \multirow{2}{*}{\textbf{TTS from No-F0 VC}} &
  \multirow{2}{*}{\textbf{TTS from F0 VC}} &
  \multirow{2}{*}{\textbf{p-value}} &
  \multirow{2}{*}{\textbf{\% improvement}} \\
   &
   &
   &
   &
   \\
  \hline
\textbf{Canadian French} & 
\textbf{Female} & 
72.09±1.39 & 
68.27±1.35 & 
67.28±1.38 & 
0.1801 & 
/                           \\
\textbf{French}         & 
\textbf{Female} & 
68.2±0.94  & 
61.27±0.94 & 
62.03±0.94 & 
0.0953 & 
/                           \\
\textbf{}                & 
\textbf{Male}   & 
68.67±0.89 & 
56.51±0.96 & 
59.14±0.93 & 
\textbf{0.0}      & 
21.70\% \\
\textbf{Italian}         & 
\textbf{Female} & 
79.84±0.74 & 
61.96±0.85 & 
63.35±0.86 & 
\textbf{0.0001} & 
7.80\%  \\
\textbf{}                & \textbf{Male}   & 79.01±0.73 & 
61.97±0.84 & 
62.16±0.84 & 
0.5791 & 
/                           \\
\textbf{German}          & 
\textbf{Female} & 
75.47±0.82 & 
63.50±0.83 & 
62.82±0.84 & 
0.1118 & 
/                           \\
\textbf{}                & 
\textbf{Male}   & 
75.32±0.81 & 
57.32±0.96 & 
62.23±0.83 & 
\textbf{0.0}      & 
27.30\% \\
\textbf{Spanish}         & \textbf{Female} & 75.49±0.79 & 
67.22±0.87 & 
68.45±0.86 & 
\textbf{0.0028} & 
17.70\% \\
\textbf{}                & 
\textbf{Male}   & 
79.62±0.71 & 
62.25±0.9  & 
63.73±0.88 & 
\textbf{0.0001} & 
8.50\% \\
\hline
\end{tabular}
\end{table*}

\textbf{F0 distribution in the synthetic data.}
In this experiment we show that our VC technology 
generates synthetic samples that reflect the speaking style of the target speaker. We analyse the distributions of F0 and delta-F0 values of both F0-conditioned and non F0-conditioned VC models. We obtain the delta-F0 values by computing the first derivative on the F0 trajectories. Delta-F0 encodes the speed of change which we believe is an important feature for characterising styles.
To perform this analysis, we generate synthetic data with three data augmentation techniques: 1) \emph{No-F0 VC} is Copycat as in the baseline \cite{huybrechts2021low}, without F0 conditioning; 2) \emph{Orig-F0 VC} is the proposed VC model (Section \ref{subsec:vc}), where at inference time the decoder is conditioned with the original F0 of the supporting speaker, rescaled by the mean of the target speaker; 3) \emph{Pred-F0 VC} is \emph{Orig-F0 VC}, where at inference time the decoder is conditioned with the F0 trajectory produced by the F0 predictor of Section \ref{subsec:f0_pred}. 

We compare the F0 distributions of the synthetic samples: firstly, we convert the F0 values into probability distributions;
then we compute the KLD of each system with respect to the target speaker F0 (and delta-F0) distributions. 
We perform this analysis over 4 speakers from 4 locales (Canadian French, French, Italian, Spanish). We chose these speakers since they have a wide range of absolute distance in terms of F0 mean and variance with respect to their respective supporting speakers. Per-speaker F0 mean and variance have been computed across the overall speaker data in the linear frequency scaling. We observed that if a target speaker has a much higher F0 variance with respect to the supporting speaker, i.e. the target speaker is more expressive than the supporting one, it is usually harder to produce synthetic data as expressive as the target voice. Having the control over F0 can help to mitigate this effect.

In Table \ref{table:f0_distribution_analysis} we observe that, for Speaker 1, \emph{Pred-F0 VC} has the smallest KLD value with respect to the real data, while for Speaker 4, the speaker which is the closest to her supporting counter-part, the best system is \emph{Orig-F0 VC}. Generally, \emph{Pred-F0 VC} gets small KLD values for the delta-F0 distributions. Our intuition is that when the supporting and the target speakers are close in terms of F0 behaviour, using the supporting speaker original F0 (rescaled by the target speaker F0 mean) is enough to resemble the F0 distribution of the target speaker (see Speaker 4). When the speakers are further apart, as in the case of Speaker 1, having an F0-predictor can have a high impact. On the other hand, \emph{Orig-F0 VC} is also the model that leads to higher fluctuations in terms of KLD scores, which indicates that its performances are less controllable, since we do not put any restrictions on the supporting speaker except for being of the same locale and gender of the target voice. \emph{No-F0 VC} can also lead to low KLD values, however, as in the case of \emph{Orig-F0 VC}, it can be unpredictable. Ultimately, \emph{Pred-F0 VC} gives us the advantage of controllability and shows consistently low KLD values across speakers. We believe that controlling the F0 generation with an high-quality F0 predictor is a key factor in order to consistently produce synthetic data that resemble the target voice's typical prosody and speaking patterns. 

\subsection{Perceptual evaluations}
\label{subsec:TTS_experiments}
We design the experiments in order to highlight how our technology: 1) scales across languages; 2) scales across voices; 3) outperforms our baseline, the state-of-the-art data augmentation approach which does not make use of F0 \cite{huybrechts2021low}. 
A few additional observations: for space reasons, we did not provide results for the TTS systems augmented with data from \emph{Orig-F0 VC} (the VC system which uses the supporting speaker rescaled F0 to condition the decoder), since they were generally worse than the proposed approach; we also did not use a multi-speaker model as a baseline since \cite{huybrechts2021low} already showed that data augmentation leads to significant improvements. Our experiments confirmed the finding.

\textbf{Evaluation protocol.}
We evaluate our models running MUSHRA (MUltiple Stimuli with Hidden Reference and Anchor)-like \cite{series2014method} tests on Naturalness. In the evaluations we omit the reference system, while we use the original recordings, as upper anchor, and the baseline system based on \cite{huybrechts2021low}. Each one of the 200 test-cases is evaluated by 15 different users. We apply a mechanism to get the fairest possible results: we select the test utterances to be mainly long utterances, i.e. more than 95\% of the utterances contain 10 or more words. This gives the listeners a higher chance of detecting bad TTS samples. The issue with short utterances is that there is less chance of detecting incorrect intonation, rhythm, etc.

\textbf{Results.}
In Table \ref{table:mushra} we show the results for the MUSHRA Naturalness evaluations for 9 synthetic voices. We can observe that our proposed data augmentation approach, \emph{TTS from F0 VC}, statistically improves the baseline, \emph{TTS from no-F0 VC}, for 5 out of 9 voices. In the remaining 4 cases, our approach is on par with the baseline. On average, when there is an improvement, our models close the gap between the baselines and the recordings by 16.6\%. For two of the voices where we are on par with the baseline (the Canadian French and French female voices), the proposed approach is very close to recordings, being respectively at 94.5\% and 90.5\% of Naturalness relative to recordings, making it hard to beat the already high performing baseline system. These results indicate that, generally, having the control on F0 during the data augmentation phase leads to more natural TTS voices. Additionally, we demonstrate that we are able to get high-quality voices across female and male speakers belonging to 5 different locales. On average, the proposed method scores 85.92\% in terms of Naturalness with respect to natural recordings.

\section{Conclusion and future work}
In this work we present a language-agnostic methodology to build low-resource TTS conversational voices. The proposed technology augments the low-resource target data via a VC system that enables F0 controllability. An external F0 predictor ensures that the augmented samples resemble the conversational style of the target speaker. Finally, a TTS model consumes the original and augmented data. The ability of learning and controlling a speaker expressiveness means that we do not need to rely on any other conversational supporting data to build a TTS voice. Therefore our technology is largely scalable to voices and languages. Our experiments demonstrate that: 1) the F0-predictor model produces reliable F0 trajectories; 2) we are able to achieve a high level of F0 controllability in VC; 3) the voice-converted samples resemble the target speaker F0 distribution; 4) we produce high-quality TTS voices from low-resource data, being on par or improving the previous best augmented system; 5) our technology is language-agnostic, hence largely scalable.


\bibliographystyle{IEEEbib}
\bibliography{references}

\end{document}